\documentclass[aps,prb,twocolumn,showpacs]{revtex4}

\usepackage{amsmath,amssymb}
\usepackage{graphicx,epsfig}

\newcommand{\nn}{\nonumber}

\newcommand{\bk}{\mathbf{k}}

\newcommand{\pdag}{{\phantom{\dagger}}}

\newcommand{\PRL}{Phys. Rev. Lett. }

\begin{document}

\title{Energy-resolved inelastic electron scattering off a magnetic impurity}
\author{Markus Garst$^{(a)}$, Peter W\"olfle$^{(b)}$, L\'aszl\'o Borda$^{(c)}$,
Jan von Delft$^{(d)}$, and Leonid Glazman$^{(a)}$}
\affiliation{
$^{(a)}$William I. Fine Theoretical Physics Institute, University of Minnesota, Minneapolis, MN 55455, USA\\
$^{(b)}$Institut f\"ur Theorie der Kondensierten Materie, Universit\"at Karlsruhe, 76128 Karlsruhe, Germany\\
$^{(c)}$ Budapest University of Technology and Economics, Institute of Physics
  and Research Group 'Theory of Condensed Matter' of the 
Hungarian Academy of Sciences H-1521 Budapest, Hungary\\
$^{(d)}$ Physics Department, Arnold Sommerfeld Center for Theoretical
  Physics, and Center for NanoScience,
  Ludwig-Maximilians-Universit\"at M\"unchen, 80333 M\"unchen,
  Germany}
\date{\today}
\begin{abstract} 
  We study inelastic scattering of energetic electrons off a Kondo
  impurity. If the energy $E$ of the incoming electron (measured from
  the Fermi level) exceeds significantly the Kondo temperature $T_K$,
  then the differential inelastic cross-section $\sigma (E,\omega)$,
  {\it i.e.}, the cross-section characterizing scattering of an
  electron with a given energy transfer $\omega$, is well-defined. We
  show that $\sigma (E,\omega)$ factorizes into two parts.  The
  $E$--dependence of $\sigma (E,\omega)$ is logarithmically weak and
  is due to the Kondo renormalization of the effective coupling. We
  are able to relate the $\omega$--dependence to the spin-spin
  correlation function of the magnetic impurity. Using this relation,
  we demonstrate that in the absence of magnetic field the dynamics of the
  impurity spin causes the electron
  scattering to be inelastic at any temperature. At temperatures $T$ low
  compared to the Kondo temperature $T_K$, the cross-section is
  strongly asymmetric in $\omega$ and has a well-pronounced maximum at
  $\hbar\omega\sim T_K$. At $T\gg T_K$, the dependence $\sigma$ {\sl
    vs.} $\omega$ has a maximum at $\omega=0$; the width of the
  maximum exceeds $T_K/\hbar$ and is determined by the Korringa
  relaxation time of the magnetic impurity. Quenching of the spin
  dynamics by an applied magnetic field results in a finite elastic
  component of the electron scattering cross-section.  The
  differential scattering cross-section may be extracted
  from the measurements of relaxation of hot electrons injected in
  conductors containing localized spins.
\end{abstract}
\maketitle
\section{Introduction}
Scattering of an electron off a magnetic impurity embedded in a
conductor is known to be anomalously strong~\cite{Abrikosov-textbook}.
The origin of the anomaly is rooted in the degeneracy of the localized spin
states. This degeneracy, being removed by a weak exchange interaction
with the itinerant electrons in a metal, gives rise to the strong
scattering of electrons with low energy -- the Kondo effect.
Perturbation theory in the exchange interaction constant $J$ is
singular. The second-order contribution in $J$ to the scattering
amplitude diverges logarithmically~\cite{Kondo} if the electron energy
$E$ (measured from the Fermi level) and temperature $T$ are
approaching zero. It is important to notice that the logarithmically
divergent contribution to the amplitude corresponds to an elastic
process. Indeed, this contribution comes from the change of state of
{\em one} electron; states of all other itinerant electrons are the
same in the beginning and end of the scattering process. Therefore,
the energies of the electron before and after the scattering is unchanged. 

The divergence noticed by Kondo is not unique to the second order of the
perturbation theory. Its higher orders ($n>2$) also contain divergent terms of
the type $J^n\ln^{n-1}(D/\varepsilon)$, where $\varepsilon={\rm max}(E, T)$,
and $D$ is some ultraviolet energy cut-off, whose value depends on the specific
model; $D\gg \varepsilon$. These leading logarithmic terms may be summed up by
diagrammatic method~\cite{Abrikosov1965} or by means of the ``Poor man's
scaling''~\cite{Anderson} renormalization group (RG), yielding for the
scattering amplitude
\begin{equation}
A_{k,\sigma,S\to k',\sigma', S'}=\frac{1}{\ln(\varepsilon/T_K)}{\bf
  s}_{\sigma,\sigma'}\cdot{\bf S}_{s,s'}.
\label{eq:Amplitude}
\end{equation}
where ${\bf s}_{\sigma,\sigma'}$ and ${\bf S}_{s,s'}$ are the spin operators
of the conduction electrons and the impurity, respectively.  The so-called
Kondo temperature is given in terms of the cut-off, $D$, and the exchange
interaction, $J$, as $T_K = D e^{-1/(J \nu)}$, where $\nu$ is the density of
states.  Like the lowest-order perturbation theory result, the
leading-logarithmic approximation Eq.~(\ref{eq:Amplitude}) corresponds to
purely elastic electron scattering.

The leading-logarithmic approximation is adequate at $\varepsilon\gg T_K$, but
it fails at low temperatures. A convenient phenomenological
description of the low-energy behavior of a single-channel Kondo model
is given by Nozi\`eres' effective Fermi liquid theory. In this
theory, a scattering problem can be formulated, too. It is clearly
seen~\cite{Nozieres}, however, that the scattering is not purely
elastic at $\varepsilon \ll T_K$. At $T=0$, for example, the inelastic
contribution to the electron scattering cross-section scales as
$(E/T_K)^2$, and becomes comparable to the elastic part at
$E\sim T_K$.

The Kondo effect is a crossover phenomenon, rather than a phase
transition. The measurable characteristics, such as the contribution
to the susceptibility or resistivity due to magnetic impurities depend
smoothly on temperature. Similarly, the electron scattering off a
magnetic impurity, which is deeply inelastic at $\varepsilon\sim T_K$,
must have some inelastic component at any energy $E$. In this paper we
investigate in detail the inelastic scattering of a high-energy
electron off a magnetic impurity.

A study of the energy-resolved, differential cross-section, $\sigma
(E,\omega)$, is interesting in its own right, but it can, in principle, also
be measured, {\it e.g.}, in a modification of the experiments of Pothier {\sl
et al.}\cite{Pothier}.  Further motivation to study $\sigma (E,\omega)$ beyond
perturbation theory comes from the recent theoretical work of Zar\'and {\it et
al.}~\cite{Natan}. In Ref.~\onlinecite{Natan} the energy dependence of the
{\em total} scattering cross-section, $\sigma_{\rm tot}(E) = \int d\omega\,
\sigma(E,\omega) $, was addressed. With the help of the optical theorem, the
total cross-section $\sigma_{\rm tot}(E)$ was compared with the elastic part
of it.  The conclusion reached in Ref.~\onlinecite{Natan} regarding the energy
domain $E\gg T_K$ is striking: at $T=0$ the scattering is deeply inelastic;
the elastic part turns out to be negligibly small. This seemingly contradicts
the leading-logarithmic result for the scattering amplitude
Eq.~(\ref{eq:Amplitude}). The physical explanation of this phenomenon however
remained unclear in Ref.~\onlinecite{Natan} and motivates us to revisit the
problem of inelastic scattering. The dependence of the differential
cross-section $\sigma(E,\omega)$ on $\omega$, which we consider in this paper,
clarifies the issue, as we are able to determine the distribution of energy
losses in the inelastic electron scattering off a magnetic impurity.

The separation of the electron scattering cross-section in the Kondo effect
into elastic and inelastic parts at $E \gg T_K$ was not addressed for decades,
as it does not affect the routinely measured quantity, the resistivity. The
anomalously fast electron energy relaxation in some mesoscopic metallic
wires~\cite{Pothier}, which was discovered in the last decade, prompted a
search for relaxation mechanisms driven by impurities with internal degrees of
freedom. A viable mechanism of energy relaxation was suggested first in
Ref.~\onlinecite{Kaminski}, and was associated with the electron-electron
scattering mediated by exchange interaction of electrons with magnetic
impurities. The removal of degeneracy of the localized spin states by the
exchange interaction results in an anomaly of the electron-electron scattering
cross-section at small energy transfers~\cite{Kaminski}; the collision of two
electrons with energies $E,E'\gg T_K$ leads to a re-distribution of the
energies between the two particles, $E,E'\to E-\omega,E'+\omega$, and has
cross-section $K(\omega,E,E')\propto J^4/\omega^2$ in the lowest-order
perturbation theory~\cite{zavsol}. The $1/\omega^2$ dependence of $K$ allowed
the experimental observations~\cite{Pothier} to be explained
qualitatively. Later experiments~\cite{Anthore} performed in a magnetic field
sufficient for the Zeeman splitting of impurity energy levels did confirm the
origin~\cite{Kaminski} of the inelastic electron-electron scattering, and
indicated the irrelevance of more exotic mechanisms, which assumed a generic
non-Fermi liquid behavior introduced by impurities~\cite{Kroha}.

The existence of energy exchange between electrons mediated by
their interaction with a magnetic impurity indicates the inelastic
nature of the electron scattering off a magnetic impurity. Indeed,
using the Fermi Golden rule we find
\begin{align}
\sigma (E,\omega) 
\propto
\frac{J^4}{\omega^2}
\int_{-\infty}^{\infty} dE' f(E')(1-f(E'+\omega))
\propto \frac{J^4}{\omega}
\label{inelast}
\end{align}
at $\omega\gg T$. So already in the simplest perturbation theory it becomes
clear that there is an inelastic contribution to scattering.  As long as $E,
\omega \gg T$, temperature does not affect the inelastic cross-section in this
order. It is not clear, however, what the relation is between the inelastic
cross-section $\sigma (E,\omega)$ and the leading-logarithmic result
Eq.~(\ref{eq:Amplitude}): On one hand, $\sigma (E,\omega)\propto J^4$ is
parametrically smaller than the scattering cross-section following from
Eq.~(\ref{eq:Amplitude}).  On the other hand, the total inelastic
cross-section obtained from Eq.~(\ref{inelast}), $\sigma_{\rm tot}(E)=\int
d\omega\,\sigma (E,\omega)$, diverges at $\omega\to 0$ indicating the
inapplicability of the lowest-order perturbation results at small energy
transfers.
  
The lowest-order perturbation theory for $K(\omega,E,E')$ can be controllably
improved in two respects. First, at $E,E'\gg T_K$ and $|\omega|\ll E, E'$ the
four constants $J$ entering as a product in the perturbative result, may be
replaced~\cite{Kaminski} by the properly renormalized~\cite{Anderson}
quantities~\cite{kroha1}.  Second, the divergence at $\omega\to 0$ is cut off
due to the dynamics of localized spin. An adequate theory may be developed for
high temperatures, $T\gg T_K$, where the cut-off occurs due to the Korringa
relaxation~\cite{Kaminski}. These improvements allow one to see that
$\sigma_{\rm tot}(E)$ is finite, but are insufficient to investigate the
details of the $\sigma (E,\omega)$ dependence.

In this paper we concentrate on the differential cross-section,
$\sigma(E,\omega)$, of inelastic scattering of a highly excited
electron with energy $E \gg T_K$. Despite the many-body nature of the Kondo
effect, this quantity is well-defined at $\omega \ll E$. We show
in Section~\ref{sec:Relation} that in the limit $E\gg T_K$ the
differential cross-section is related to the dissipative part of the
impurity spin susceptibility, $\chi''$. From the low- and
high-frequency asymptotes of $\chi''$ we extract in
Sections~\ref{sec:WithoutZeeman} and \ref{sec:WithZeeman} the behavior
of the differential cross-section $\sigma(E,\omega)$ in the absence
and presence of a Zeeman energy, respectively. The analytical
  asymptotes thus obtained are complemented by results of the
  numerical renormalization group\cite{Wilson} (NRG), which allows us to access
  also the intermediate range of frequencies and magnetic fields. The
connection between the result of Ref.~\onlinecite{Natan} and the
leading-logarithmic approximation for the scattering amplitude
(\ref{eq:Amplitude}) describing only elastic scattering will be
explained in detail. Finally, in Section~\ref{sec:Experiments} we
discuss possible hot-electron experiments in metallic mesoscopic wires
and in a semiconductor quantum-dot setup in order to measure the
differential scattering cross-section $\sigma(E,\omega)$.

\section{Relation between inelastic scattering cross-section and susceptibility}
\label{sec:Relation}

The relation between the scattering cross-section of a ``foreign''
spin-carrying particle and the spin-spin correlation function of a magnetic
medium is well-known from the theory of neutron
scattering~\cite{MagnetismReference}. Here we derive a similar relation for scattering
off a magnetic impurity of a high-energy electron belonging itself to the
Fermi liquid hosting the magnetic impurity.

The exchange interaction between the impurity spin and spins of
electrons forming the Fermi liquid,
\begin{equation}
H_{\rm int} = J\sum_{\bk,\bk'}{\bf S}\cdot{\bf s}_{\alpha\alpha'}
              c^\dag_{\bk \alpha} c^\pdag_{\bk' \alpha'}
\label{eq:1}
\end{equation}
gives rise to the Kondo effect. Here $J$ is the constant of exchange
interaction between the impurity spin and itinerant electrons with energies $\epsilon_{\bk \sigma}$
(measured from the Fermi level) confined to some energy band, $|\epsilon_{\bk
\sigma}|<D$. Here ${\bf s}_{\alpha\alpha'}$ is $\frac{1}{2}$ times the vector of Pauli matrices.
The Kondo problem allows for a logarithmic renormalization: the
low-energy properties of the system described by the Hamiltonian (\ref{eq:1})
coincide with those for a Hamiltonian defined in a narrower band, say
$|\epsilon_{\bk\sigma}|< E$, upon the proper renormalization of the exchange
constant,
\begin{equation}
J(E)=\frac{J(D)}{1-\nu J(D)\ln (D/E)},\quad J(D)=J,
\label{eq:2}
\end{equation}
where $\nu$ is the density of states of itinerant electrons. The perturbative
renormalization Eq.~(\ref{eq:2}) is valid as long as the running energy [$E$
in the case of Eq.~(\ref{eq:2})] significantly exceeds the Kondo temperature
$T_K$. An important property of the logarithmic renormalization is that only
exponentially wide energy intervals $(\varepsilon_1,\varepsilon_2)$, such that
$\nu J|\ln(\varepsilon_1/\varepsilon_2)|\sim 1$ contribute significantly to
the renormalization. That allows us to ``skip'' some relatively narrow strip
of energies, say, $(E-\Delta E, E+ \Delta E)$, with $\Delta E \ll E$, in the
renormalization process, yielding a Hamiltonian
\begin{align}\label{eq:3}
H_{\rm int} &=
J(\tilde{D})\sum_{|\epsilon_{\bk\alpha}|,|\epsilon_{\bk'\alpha'}|<\tilde{D}}
              {\bf S}\cdot{\bf s}_{\alpha\alpha'}
              c^\dag_{\bk \alpha} c^\pdag_{\bk' \alpha'}
\nonumber\\
&+
J(E)
\hspace{-1em}
\sum_{E-\Delta E<\epsilon_{\bk \alpha},\epsilon_{\bk'\alpha'}<E+ \Delta E}
\hspace{-1em}
              {\bf S}\cdot{\bf s}_{\alpha\alpha'}
              c^\dag_{\bk \alpha} c^\pdag_{\bk' \alpha'}
\\\nonumber
&+
J(E)
\hspace{-2em}
\sum_{
\begin{array}{c}
 \scriptstyle  
E-\Delta E<\epsilon_{\bk\alpha}<E+ \Delta E,|\epsilon_{\bk'\alpha'}|<\tilde{D};\\
\scriptstyle
E-\Delta E<\epsilon_{\bk'\alpha'}<E+ \Delta E,|\epsilon_{\bk\alpha}|<\tilde{D}
\end{array}}
\hspace{-2em}
              {\bf S}\cdot{\bf s}_{\alpha\alpha'}
              c^\dag_{\bk \alpha} c^\pdag_{\bk' \alpha'},
\end{align}
with $\tilde{D}\leq E-\Delta E$. The renormalized exchange constants here may be
expressed in terms of the Kondo temperature, $\nu
J(\varepsilon)=1/\ln(\varepsilon/T_K)$. There is no need to distinguish
between $J(E-\Delta E)$, $J(E)$ or $J(E+\Delta E)$ as long as $E \gg T_K$.

If the scattering of an electron with initial energy $E$ leaves it
in the energy domain $(E-\Delta E, E+ \Delta E)$, then the
corresponding cross-section, within the lowest-order perturbation
theory in $J(E)$, can be evaluated with the help of the
Hamiltonian (\ref{eq:3}). The first line of Eq.~(\ref{eq:3}) plays the
role of the Hamiltonian of a magnetic medium in the neutron scattering
problem, and the second line describes the interaction of the
energetic particle (we deal with an electron rather than with a
neutron though) with the medium. The remaining part of the Hamiltonian
does not contribute to the scattering cross-section in the
lowest-order calculation.

Consider such a scattering of an energetic electron with energy $E$ and spin
$\sigma$ in the initial and $E-\omega$ and $\sigma'$, respectively, in the
final state with $\omega \ll E$ such that $E-\omega \in (E-\Delta E, E+ \Delta E)$. The
state of the remaining system before and after scattering may be characterized
by the wave functions $\Psi_i$ and $\Psi_f$, respectively. The initial and
final state of the total system is then given by the product states
\begin{align}
\begin{array}{ll}
\mbox{initial state:} & 
|i\rangle =  |E,\, \sigma \rangle \otimes | \Psi_i \rangle 
\vspace{0.5em}\\
\mbox{final state:} & 
|f\rangle =  |E-\omega,\, \sigma' \rangle \otimes | \Psi_f \rangle \,.
\end{array}
\end{align}

The differential cross-section of inelastic scattering
$\sigma_{\sigma'\sigma}(E,\omega)$ is determined
by the probability $P_{\sigma'\sigma}(E,\omega) d\omega$ of scattering
of an electron with initial energy $E$ and spin $\sigma$ into a state
within interval of energies $(E-\omega, E-\omega-d\omega)$ and spin
$\sigma'$,
\begin{align} \label{DefDiffCrossSection}
P_{\sigma'\sigma}(E,\omega)\, d\omega=
v_F \sigma_{\sigma'\sigma}(E,\omega)d\omega\,,
\end{align}
where $v_F$ denotes the Fermi velocity.
By energy conservation, $\omega=\xi_f-\xi_i$, where energies $\xi_i$, $\xi_f$
are associated respectively with the functions $\Psi_i$ and $\Psi_f$ involving
the states in the domain $|\epsilon_{\bk \sigma}|<\tilde{D}$. In the absence of a
magnetic field, energy $E$ is the orbital energy in the initial state, and
$\omega$ is the change in the orbital energy resulting from scattering.  In
the presence of Zeeman splitting, the initial energy and the energy transfer
include the orbital and Zeeman parts, e.g.~$E= \epsilon_{\bk} + \sigma g_e \mu_B
B/2$.

The standard application of the lowest-order
perturbation theory in the interaction of the energetic electron with
the remaining system yields for the scattering probability
\[
w_{f \leftarrow i} = 
|J(E){\bf s}_{\sigma\sigma'} \langle\Psi_f|{\bf S}|\Psi_i\rangle|^2
2\pi\nu\delta(\xi_i - \xi_f + \omega),
\]
where $\nu$ is the density of states for the energetic ($\epsilon\sim E$)
electron.  After the summation over the final states and proper thermal
averaging over the initial states, we are able to relate $w_{f \leftarrow i}$
with $\sigma_{\sigma'\sigma}(E,\omega)$, and obtain the differential
scattering cross-section
\begin{align}  \label{eq:4}
\lefteqn{\sigma_{\sigma'\sigma}(E,\omega)
=\frac{\nu}{4 v_F} J^2(E)}
\\\nn
&\qquad\qquad\times
\left(\delta_{\sigma' \sigma} \mathcal{S}_{zz}(\omega) + 
{\bf s}^+_{\sigma' \sigma}
\mathcal{S}_{+-}(\omega)  + 
{\bf s}^-_{\sigma' \sigma}
\mathcal{S}_{-+}(\omega)
\right),
\end{align}
where ${\bf s}^\pm_{\sigma' \sigma} = {\bf s}^x_{\sigma' \sigma} \pm i
{\bf s}^y_{\sigma' \sigma}$.  As in the theory of neutron
scattering~\cite{MagnetismReference}, the cross-section involves a
spin-spin correlation function. Here it is the correlation function of
the local magnetic impurity spin,
\begin{align} \label{SpinCorrelator}
\lefteqn{
\mathcal{S}_{ab}(\omega) = \int\limits_{-\infty}^\infty dt\, e^{i \omega t} 
\langle S^a(t) S^b(0) \rangle }
\\\nn&=
\!\!\!\!\!\sum_{\{|\Psi_i\rangle\,,|\Psi_f\rangle\}} \!\!\!\!\! 
\frac{e^{-\beta\xi_i}}{Z}
\langle \Psi_i | S^a | \Psi_f \rangle 
\langle \Psi_f | S^b | \Psi_i \rangle
2 \pi \delta(\xi_i - \xi_f +\omega)
\,.
\end{align}
We thus reduced the scattering cross-section to an expression where
its dependence on the energy of the scattering hot electron, $E$,
separates from the dependence on the energy loss $\omega$. The
dependence on the energy loss is determined by the dynamics of the
impurity spin characterized by the correlation function $\mathcal{S}$.
The spin correlator is related to the dissipative part of the impurity
susceptibility via the fluctuation-dissipation theorem,
\begin{align}\label{eq:10}
(g\mu_B)^2\mathcal{S}_{ab}(\omega) = \frac{2}{1-e^{-\beta \omega}}
\chi''_{ab}(\omega)\,.
\end{align}
Here $\mu_B$ is the Bohr magneton, and $g$ is the impurity $g$-factor.  The
behavior of $\chi''$ in various limits will be discussed in the following
sections.  The spin dynamics is thus included in a non-perturbative
fashion. It will allow us to investigate the behavior of the cross-section at
any energy transfer; at $\omega\ll T_K$ we apply effective Fermi liquid
theory, and the region of intermediate energies, $\omega\sim T_K$, is covered
with the help of NRG calculations.

However, it is important to note that the total scattering cross-section is
fixed by the sum rule for the spin correlation function,
$\mathcal{S}_{ab}(\omega)$.  Consider the total cross-section obtained after
averaging over the initial electronic spin configurations, $\sigma$, summing
over the final ones, $\sigma'$, and integrating over the energy transfer
$\omega$,
\begin{align} \label{TotalCrossSection}
\sigma_{\rm tot}(E) = \frac{1}{2} \sum_{\sigma,\sigma'} \int\limits_{-\infty}^\infty d\omega\, \sigma_{\sigma'\sigma}(E,\omega) 
=
\frac{3 \pi}{8} \frac{1}{\nu\,v_F} \frac{1}{\ln^2 \frac{E}{T_K}}.
\end{align}
We substituted the explicit form for the energy dependent exchange
interaction, $J(E) = 1/(\nu \ln(E/T_K))$. The total scattering cross-section
will be used throughout the rest of the paper as a convenient basic unit of
measurement for the differential cross-section discussed below.

As we are mainly interested in the dependence of the scattering probability on
the energy transfer, $\omega$, we will confine ourselves in the following to
an analysis of the scattering cross-section averaged over the initial
electronic spin configurations, $\sigma$, and summed over the final ones,
$\sigma'$,
\begin{align} \label{CrossSectionSpinAveraged}
\sigma(E,\omega) =  \sigma_{\rm tot}(E) \frac{2}{3\pi}
\left[\mathcal{S}_{zz}(\omega) + \frac{1}{2} 
\left(\mathcal{S}_{+-}(\omega) + \mathcal{S}_{-+}(\omega)\right)\right].
\end{align}
Note that a Zeeman energy of electrons forming the Fermi sea was already
incorporated in the definition of the energies $E$ and $\omega$.  The
generalization of our results to spin-resolved scattering is straightforward.

\section{Inelastic electron scattering in the absence of Zeeman
  splitting}
\label{sec:WithoutZeeman}

In the absence of a magnetic field the expression for the scattering
cross-section (\ref{CrossSectionSpinAveraged}) simplifies considerably since
the impurity spin correlator is diagonal, $\mathcal{S}(\omega) \equiv
\mathcal{S}_{zz}(\omega) = \frac{1}{2}\mathcal{S}_{+-}(\omega)$,
%
\begin{align}   \label{CrossSectionWithoutB}
\sigma(E,\omega) = \sigma_{\rm tot}(E)\,
\frac{2}{\pi} \mathcal{S}(\omega) \,.
\end{align}
Let us first establish the relation between
Eq.~(\ref{CrossSectionWithoutB}) and the well-known result of the
leading-logarithmic approximation~\cite{Abrikosov1965,Anderson}. For
that, we need to substitute in Eq.~(\ref{CrossSectionWithoutB})
the function $\mathcal{S}(\omega)$ evaluated in the zeroth order in the
exchange interaction $J(\tilde{D})$. In this order 
$\mathcal{S}^{(0)}(\omega) = \frac{\pi}{2} \delta(\omega)$, which
yields the well-known result~\cite{Abrikosov1965,Anderson} for the
cross-section, 
\begin{align} \label{AbrikosovScattering} 
\sigma^{(0)}(E,\omega) = \sigma_{\rm tot}(E)\, \delta(\omega)\,,
\end{align}
{\em i.e.}, scattering is elastic in the leading-logarithmic
approximation. The elasticity breaks down, however, if one accounts for
$J({\tilde D})\neq 0$. Indeed, the exchange interaction $J(\tilde{D})$ leads
to some dynamics of the impurity spin. The delta-function 
in 
Eq.~(\ref{AbrikosovScattering}) gets broadened, and spectral weight is
transfered to finite energies $\omega\neq 0$. The shape of the broadened peak
is related to the character of the spin dynamics, which is different in the
limits of high, $T\gg T_K$, and low, $T\ll T_K$ temperatures. We study the
shape of the peak in these limits below. However, note that the
broadening does not affect the {\em total} cross-section
which is fixed by the sum rule 
and remains the same as for the elastic
scattering, Eq.~(\ref{AbrikosovScattering}), evaluated in the leading
logarithmic approximation.

\subsection{Inelastic electron scattering at $T \gg T_K$}

At $T\gg T_K$, the local spin exhibits relaxational dynamics. The
Bloch equations for the impurity spin in the absence of a magnetic field,
\begin{align}
\frac{\partial }{\partial t} \langle S^a \rangle 
= - \frac{1}{\tau_K} \langle S^a \rangle\,,
\end{align}
imply the following form for the imaginary part of the
susceptibility\cite{Goetze71}, $\chi''_{ab}(\omega) = \delta_{ab}
\chi''(\omega)$ with
\begin{align} \label{BlochSusceptibility}
\chi''(\omega) = \chi_0(T) \frac{\omega /\tau_K}{\omega^2 + (1/\tau_K)^2}\,.
\end{align}
It involves the static susceptibility which is given by $\chi_0(T) =
(g\mu_B)^2/(4 T)$. The decay time, $\tau_K$, in the Bloch equations is
the Korringa relaxation time~\cite{Korringa}, $1/\tau_K = \pi (\nu
J(T))^2 T$. Inserting the scale dependent exchange interaction,
$J(T)$, the Korringa relaxation rate reads explicitely
\begin{align}\label{eq:15}
\frac{1}{\tau_K} = \frac{\pi T}{\ln^2\frac{T}{T_K}}\,.
\end{align}
It is parametrically smaller than $T$ at temperatures $T\gg T_K$.  

Expression (\ref{BlochSusceptibility}) adequately accounts for the behavior of
$\chi''$ at low frequencies, $\omega \lesssim T$, but fails at higher
frequencies.  For $\omega \gg T$, the susceptibility can be
evaluated within the lowest-order perturbation theory in the
exchange constant\cite{Koller}, $J({\tilde D})$,
\begin{align}
(g\mu_B)^{-2}\chi''(\omega) = \frac{\pi}{4} \frac{1}{\omega
  \ln^2\frac{|\omega|}{T_K}}\,.
\label{eq:15a}
\end{align}
The additional logarithmic frequency dependence arises from the logarithmic
enhancement of the exchange interaction due to the perturbative RG, which is
now cut-off at a band width $\tilde{D} \sim \omega$.

\begin{figure}
\includegraphics[width= 0.9\linewidth]{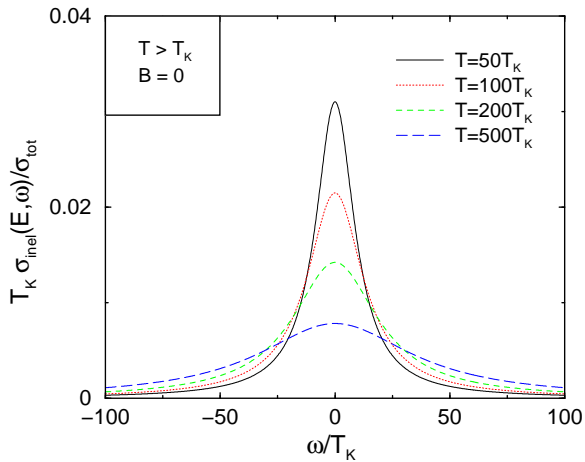}
\caption{\label{fig:1a} (Color online) Differential cross-section, $\sigma(E,\omega)$, at
large temperatures, $T \gg T_K$, without Zeeman splitting, $B=0$, as given by
Eq.~(\ref{eq:16}).  The Lorentz peaks have a width given by the Korringa
relaxation rate $1/\tau_K$, Eq.~(\ref{eq:15}).}
\end{figure}
The resulting differential cross-section $\sigma (E,\omega)$ can be found with
the help of Eq.~(\ref{eq:10}). It is symmetric in $\omega$ at small energy
transfers. It shows a narrow peak at $\omega = 0$ and falls off significantly
within the region of energies $|\omega|\lesssim T$:
\begin{equation}\label{eq:16}
\sigma(E,\omega)= \sigma_{\rm tot}(E)\,\delta_\Gamma(\omega),
\end{equation}
where we introduced a ``broadened delta-function'', which is a Lorentzian with
linewidth $\Gamma=1/\tau_K$,
\begin{equation}\label{eq:16a}
\delta_\Gamma(\omega)=\frac{1}{\pi}\frac{1/\tau_K}{\omega^2+(1/\tau_K)^2}.
\end{equation}
As $1/\tau_K\ll T$, see Eq.~(\ref{eq:15}), almost the full weight of the total
cross-section is accounted for by 
Eqs.~(\ref{eq:16}) and (\ref{eq:16a}), see Fig.~\ref{fig:1a}.

At higher energy transfers, $|\omega|\gtrsim T$, the cross-section is
asymmetric in $\omega$,
\begin{equation}
\sigma (E,\omega)
=\sigma_{\rm tot}(E)\,
\frac{1}{1-e^{-\omega/T}}
\frac{1}{\omega \ln^2(|\omega|/T_K)}.
\label{eq:17}
\end{equation}
The probability for the scattered electron to acquire energy ($\omega<0$) is
exponentially suppressed. Although the contribution of Eq.~(\ref{eq:17}) to
the total cross-section is parametrically small, $\propto 1/\ln(T/T_K)$, it
is worth noting that its decay with $\omega$ is remarkably slow.

The slow decay of $\sigma (E,\omega)$ {\it vs} $\omega$ is related to the
dependence on the transfered energy of the cross-section for inelastic
electron-electron scattering mediated by a magnetic
impurity~\cite{Kaminski}. The probability for such an inelastic scattering
between two electrons with initial energies $E$ and $E'$ and final energies
$E-\omega$ and $E'+\omega$ was calculated in Ref.~\onlinecite{Kaminski}; all
of these four energies were assumed to be large compared to $T_K$. According
to Ref.~\onlinecite{Kaminski} [see also Eq.~(9) of Ref.~\onlinecite{Ujsaghy}],
the contribution $K(\omega;E,E')$ of a single magnetic impurity to this
probability in the limit of high energy, $E\gg |\omega|$, reads
\begin{equation}
K(\omega;E,E') = \frac{3\pi}{8 \nu}
\frac{1}{\ln^2\frac{|E|}{T_K}}
\frac{4}{\left(\ln\frac{|E'|}{T_K}+\ln\frac{|E'+\omega|}{T_K}\right)^2} 
\frac{1}{\omega^2}\,.
\label{eq:19}
\end{equation}
The differential cross-section (\ref{eq:17}) can be obtained by
integrating $K(\omega;E,E')$ over the available phase space volume of
one of the scattering electrons,
\begin{align} 
v_F\,\sigma(E,\omega) = \int dE' f(E')(1-f(E'+\omega)) K(\omega;E,E').
\label{eq:18}
\end{align}
The Fermi functions in Eq.~(\ref{eq:18}) confine the energy $E'$ to an
interval $-\omega \lesssim E'\lesssim 0$. This includes a regime where
the arguments of the $E'$--dependent logarithmic factors are not
meaningful anymore and should be replaced by temperature or the Korringa
relaxation rate. After such cut-off, integration over $E'$ is easily
performed, yielding $\ln^{-2}|\omega|/T_K$ within logarithmic
accuracy. This way, starting from the collision integral kernel of
Ref.~\onlinecite{Kaminski}, one recovers Eq.~(\ref{eq:17}).

\subsection{Inelastic electron scattering at $T\ll T_K$}

When the temperature is below the Kondo temperature the picture differs
drastically from the zeroth order result (\ref{AbrikosovScattering}). For $T
\ll T_K$, the low-frequency behavior of the scattering cross-section is
beyond perturbation theory. Nevertheless, the cross-section for small
energy transfers, $|\omega| \ll T_K$, may be found with the help of the
Shiba relation~\cite{Shiba} for the susceptibility,
\begin{equation}
(g\mu_B)^2\chi''(\omega) =2\pi\omega \left[\chi_0(T=0)\right]^2\,.
\label{shiba}
\end{equation}
The zero-temperature static susceptibility $\chi_0(0)$ is used
conventionally\cite{Hewson} to define the pre-exponential factor of the Kondo
temperature, $\chi_0(0) = [(g\mu_B)^2W]/(4 T_K)$; here $W = 0.413...$ is
Wilson's number. (We present a convenient derivation of the Shiba relation in
Appendix~\ref{app:ShibaDerivation}.) The corrections to the Shiba relation are
of order $\mathcal{O}\left(\omega T^2/T_K^2, \omega^3/T_K^2\right)$ and are
sub-leading. We thus obtain for the cross-section at $|\omega|, T \ll T_K$:
\begin{equation}
\sigma(E,\omega) = \sigma_{\rm tot}(E) 
\frac{W^2}{2} \frac{1}{1-e^{-\omega/T}} \frac{\omega}{T_K^2}.
\label{eq:20}
\end{equation}
The high-frequency limit, $|\omega|\gg T_K$, of the
scattering cross-section can still be obtained perturbatively and is
given by Eq.~(\ref{eq:17}).

\begin{figure}
\includegraphics[width= 0.9\linewidth]{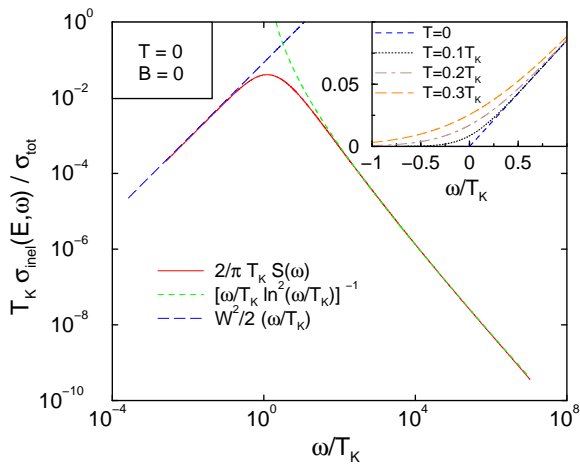}
\caption{\label{fig:1b} (Color online) NRG result for $\sigma(E,\omega)$ (solid line) on a
logarithmic scale at $T=0$ without Zeeman splitting, $B=0$. A maximum at
finite $\omega = T_K$ develops and scattering with small energy transfer
$\omega$ is suppressed. Whereas the high-frequency tail is perturbatively
accessible, see Eq.~(\ref{eq:17}) (short-dashed line), the low-frequency tail,
Eq.~(\ref{eq:20}) (long-dashed line), is a property of the strong coupling
fixed point described by Nozi\`eres' Fermi liquid theory. ($W=0.413...$ is
Wilson's number.) The inset shows the temperature correction according to
Eq.~(\ref{eq:20}); the contribution for negative $\omega$ is exponentially
small for temperatures $0 < T \ll T_K$.}
\end{figure}

Comparing the results Eq.~(\ref{eq:17}) and Eq.~(\ref{eq:20}) we see that for
temperatures $T \ll T_K$ the differential cross-section $\sigma(E,\omega)$
peaks at energy transfers of the order of $\omega\sim T_K$.  It then decreases
linearly upon further decrease of $\omega$, until it crosses over (at
$|\omega|\lesssim T$) into the exponential tail for $\omega<0$, see inset of
Fig.~\ref{fig:1b}. At zero temperature the factor containing $\exp(-\omega/T)$
in Eq.~(\ref{eq:20}) becomes a step function which forbids any energy gain
from the Kondo system,
\begin{equation}
\sigma(E,\omega) = \sigma_{\rm tot}(E) 
\frac{W^2}{2} \Theta(\omega) \frac{\omega}{T_K^2};
\label{eq:21}
\end{equation}
here $\Theta(x) = 1$ if $x>0$ and $0$ if $x<0$.

The region between the asymptotes given in Eqs.~(\ref{eq:17}) and
(\ref{eq:20}) can be bridged by calculations performed with the NRG method. In
this method, after the logarithmic discretization of the conduction band one
maps the Kondo Hamiltonian onto a semi-infinite chain with the impurity at the
end. As a consequence of the logarithmic discretization, the hopping along the
chain decreases exponentially, $t_n\sim\Lambda^{-n/2}$, where $\Lambda>1$ is
the discretization parameter and $n$ is the site index.  (We have used
$\Lambda=2$ throughout the calculations presented in the paper.) The
separation of energy scales provided by the exponential decay of the hopping
rate allows us to diagonalize the Hamiltonian iteratively and keep the
eigenstates with the lowest energy as most relevant ones. Since we know the
energy eigenvalues and eigenstates, we are able to calculate the impurity spin
correlation function directly (see Eq.(\ref{SpinCorrelator})) given that the
Dirac delta function appearing in the Lehman representation must be broadened
when performing a numerical calculation\cite{Costi,Bulla}.  The result of the
NRG calculation is shown in Fig.~\ref{fig:1b}.
  
To summarize this Section we demonstrated that the dynamics of the impurity
spin leads to inelastic electron scattering at all temperatures. The main
contribution to the total scattering cross-section comes from $\omega\sim T_K$
or $|\omega|\lesssim 1/\tau_K$ at $T\ll T_K$ and $T\gg T_K$ respectively.  The
total scattering cross-section is fixed by the sum rule for the impurity spin
correlation function, see Eq.~(\ref{TotalCrossSection}), and is thus
determined by the effective exchange constant, $J(E)$, evaluated within the
leading logarithmic approximation\cite{Abrikosov1965}.

\section{Zeeman effect in the electron scattering}
\label{sec:WithZeeman}

We now address the case when the degeneracy of the impurity spin is lifted by
a magnetic field. 
The Zeeman splitting of the impurity spin is described by the Hamiltonian
\begin{align}
H_{\rm Zeeman} = - g\mu_B {\bf S}^z B.
\label{eq:22}
\end{align}
In the presence of the Zeeman splitting the scattering electron has to pay
Zeeman energy in order to transfer spin to the Kondo system. The resonance
structure for electron scattering involving a spin-flip will therefore differ
from the one of non-spin-flip scattering.  Evaluating the impurity spin
correlator in zeroth order in $J(\tilde{D})$ we obtain for the scattering
cross-section (\ref{CrossSectionSpinAveraged}) in the leading logarithmic approximation
\begin{align}\label{eq:23}
\lefteqn{\sigma^{(0)}(E,\omega) = 
\sigma_{\rm tot}(E) 
\frac{2}{1+e^{-\beta \omega}} }
\\\nn & \qquad \qquad \times
\frac{1}{3}\,\left\{\delta(\omega) + \delta (\omega -\omega_Z(B))+ 
\delta (\omega + \omega_Z(B))\right\}
\end{align}
The single delta-function for $B=0$,
Eq.~(\ref{AbrikosovScattering}), is now split into three
contributions. In addition to a delta-function at zero frequency,
which is due to non-spin-flip scattering, there are two Zeeman
satellites at $\omega=\pm\omega_Z(B)$. In the limit of low
temperatures, $T \ll B$, the satellite at negative Zeeman energy
corresponding to an energy gain of the scattering electron is
exponentially small as it is clear from Eq.~(\ref{eq:23}).

The Zeeman energy $\omega_Z(B)$ depends on the renormalized
$g$-factor, which is different from its bare value $g$ appearing in the
Zeeman Hamiltonian (\ref{eq:22}). When we derived the
effective interaction Hamiltonian (\ref{eq:3}), we integrated out a finite
band of electronic degrees of freedom which lead to a renormalization of the
exchange interaction $J$. The Zeeman term (\ref{eq:22}) is
not invariant under this perturbative renormalization of the Kondo
model. Similar to the exchange interaction $J$, the $g$-factor is also
renormalized when the band is reduced from $D$ to $\tilde{D}$. As explained in
Appendix~\ref{app:RG-gfactor}, the scale-dependent $g$-factor in the leading
logarithmic order is given by
\begin{align} \label{eq:24}
\frac{g(\tilde{D})}{g} = 
\left( 1- \frac{1}{2 \ln \tilde{D}/T_K} \right)\,.
\end{align}
To find the observable value of $g$-factor, one needs to set
$\tilde{D}={\rm max}\{T,g\mu_BB\}$. The position of the Zeeman resonances,
to the leading logarithmic order, is given by~\cite{Moore00}
\begin{align}\label{eq:25}
\omega_Z(B) = 
g \left(1 - \frac{1}{2 \ln \left({\rm max}\{T, g\mu_B B\}/T_K\right)}
\right) \mu_B B.
\end{align}

Beyond the leading logarithmic approximation the dynamics of the local 
spin is characterized by a further redistribution of the spectral weight 
of the scattering cross-section (\ref{eq:23}). However, a striking feature of the
presence of a magnetic field is that a finite weight of the
delta-resonance at $\omega=0$ will still survive after accounting for
the coupling of the impurity spin to the low-energy degrees of freedom
of the Fermi sea. In other words, at any ratio $T/T_K$ a part of the
scattering becomes elastic if a magnetic field $B \neq 0$ is turned
on.  This can be best understood by considering the longitudinal spin
correlation function in time. For $B\neq 0$ this correlation function
will not fully decay with time but rather saturate at a value given by
the finite expectation value of the impurity spin, $\langle S^z(t)
S^z(0) \rangle \to \langle S^z\rangle^2$ for $t\to \infty$. This
finite saturation value leads to a finite weight of the delta-function
$\delta(\omega)$ in its Fourier transform and in Eq.~(\ref{eq:4}). Let
us decompose $\sigma(E,\omega)$ into the elastic and inelastic parts,
\begin{align}\label{eq:26}
\sigma(E,\omega) = \sigma_{\rm el}(E,\omega) + \sigma_{\rm inel}(E,\omega)\,.
\end{align}

The elastic part will be determined by the magnetization of the impurity spin
\begin{align} \label{eq:27}
\sigma_{\rm el}(E,\omega) = \sigma_{\rm tot}(E)\, 
\frac{4}{3} \langle S^z \rangle^2 \delta(\omega)
\,.
\end{align}
Being a thermodynamic quantity, $\langle S^z\rangle$ has a
well-studied field and temperature
dependence~\cite{bethe-ansatz,Hewson}. In the scaling regime,
$f(t,b)=\langle S^z\rangle$ is a function of $t=T/T_K$ and $b=g\mu_B
B/T_K$. The asymptote of $f(t,b)$ at ${\rm max}(t,b)\gg 1$ 
is with logarithmic accuracy given by
\begin{align}\label{eq:28}
\lefteqn{
f(t,b) =
\frac{1}{2}\left(1-\frac{1}{2\ln[{\rm max}(t,b)]}\right)
}
\\\nn& \hspace{5em} \times
\tanh\left[\frac{b}{2 t}\left(1-\frac{1}{2\ln[{\rm max}(t,b)]}\right)\right]\,.
\end{align} 
Note that in the limit $t=0$, $b\gg 1$, Eq.~(\ref{eq:28}) yields the
ground-state value of $\langle S^z\rangle$ in the perturbative regime.
In the opposite limit of a weak field, $b\ll 1\ll t$, spin polarization
is small according to the Curie law, $f\sim b/4 t$. In the developed
Kondo regime, ${\rm max}(t,b)\ll 1$, the average spin is
$f(t,b)=(W/4)b$.

\begin{figure}
\includegraphics[width= 0.9\linewidth]{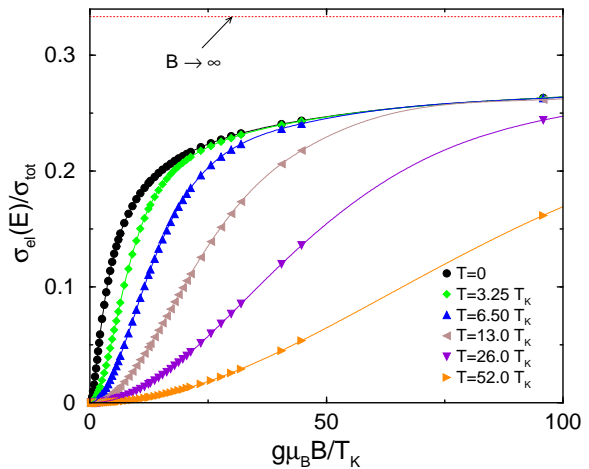}
\caption{\label{fig:3} (Color online) Weight of the elastic scattering cross-section,
  $\sigma_{\rm el}(E) = \int d\omega\, \sigma_{\rm el}(E,\omega)$, see
  Eq.~(\ref{eq:27}), determined by NRG. The weight increases as $B^2$ for
  small magnetic fields and saturates logarithmically slowly to the limiting
  value for large $B$, see Eq.~(\ref{eq:36a}).  }
\end{figure}

The weight of the elastic scattering, Eq.~(\ref{eq:27}), evaluated
with NRG is shown in Fig.~\ref{fig:3}. In the limit of small magnetic
fields this weight increases as $B^2$. The saturation of the weight to
its large-field limit, $1/3$, is remarkably slow due to the
logarithmic correction to the magnetization\cite{bethe-ansatz}, see
Eq.~(\ref{eq:28}).

The inelastic part of the scattering cross-section, $\sigma_{\rm
  inel}(E,\omega)$, accounts for the remaining spectral weight. Note
however that the total scattering cross-section, {\it i.e.} the total
spectral weight, is independent of the magnetic field; its value being
fixed by the sum rule for the impurity spin correlator.

\subsection{Dissipative part of magnetic susceptibility}

To analyze the inelastic scattering-cross section in more detail for
the two limiting cases $T \gg T_K$ and $T \ll T_K$, we start from
presenting the proper details regarding the frequency dependence of
the dissipative parts of longitudinal and transversal impurity spin
susceptibilities ($\chi''_{zz}$ and $\chi''_{+-}$, respectively).

At $T\gg T_K$, one may treat the exchange interaction $J(\tilde{D})$
perturbatively at any field $B$. The effect of $B$ on $\chi''$ is
negligible as long as the Zeeman splitting $\omega_Z(B)$ is smaller
than the Korringa relaxation rate $1/\tau_K$ , see Eq.~(\ref{eq:15}).
At higher fields, the susceptibility becomes anisotropic,
$\chi''_{zz} \neq \frac{1}{2}\chi''_{+-}$, 
and its frequency dependence acquires a well-resolved structure. The
dissipative part of the susceptibility can be found from the Bloch
equations~\cite{MagnetismReference}. The transversal part takes the form
\begin{align}\label{eq:31}
\chi''_{+-}(\omega) &=
2 \chi_T \frac{\omega/T_2}{[\omega-\omega_Z(B)]^2 + (1/T_2)^2}
 \,,
\end{align}
where the static transversal differential susceptibility can be expressed with
the help of Eq.~(\ref{eq:28}) as $\chi_T=(g \mu_B) f(t,b)/B$. The
longitudinal part reads
\begin{align}\label{eq:30}
\chi''_{zz}(\omega) &= \chi^D_L \frac{\omega/T_1}{\omega^2 + (1/T_1)^2}
 \,,
\end{align}
where $\chi^D_L$ is given by 
\begin{align}\label{StaticSuscept}
\chi^D_L &= 
\frac{(g\mu_B)^2 \left(1 - \frac{1}{\ln \left({\rm max}\{T, g \mu_B B\}/T_K\right)}\right)}
{4 T \cosh^2{\frac{\omega_Z(B)}{2 T}}}\,.
\end{align}
The factor $\chi^D_L$ can be understood as the contribution to the static suceptibility which originates from the response of the occupation factors of the two Zeeman levels to a varying magnetic field, $\chi^D_L = g_{\rm eff} \mu_B \partial \langle n_+-n_-\rangle/ \partial B$; here  $g_{\rm eff}$, see Eq.~(\ref{eq:24}), is the appropriately renormalized $g$-factor. Note that only in the limit $\omega_Z(B) \lesssim T$ when the renormalized $g$-factor (\ref{eq:24}) is insensitive to the magnetic field, $\chi^D_L$ does coincide with the full static longitudinal differential susceptibility $\chi_L = g \mu_B \partial f(t,b)/ \partial B$.

In the case of a moderately high field, $1/\tau_K\ll\omega_Z(B)\lesssim T$,
the relaxation times, $T_1$ and $T_2$, equal each other~\cite{MagnetismReference} and
are given by Eq.~(\ref{eq:15}); $T_1=T_2=\tau_K$.

At even higher fields, $\omega_Z(B)\gtrsim T$, the peak structure in
$\chi''_{+-}(\omega)$ is still described by a Lorentzian form of
Eq.~(\ref{eq:31}) but the corresponding relaxation time is determined
now by the Zeeman splitting rather than by temperature\cite{Goetze71},
\begin{align}\label{eq:32}
  \frac{1}{T_2} = \frac{\pi}{4} \frac{\omega_Z(B)}{\ln^2
    \frac{\omega_Z(B)}{T_K}}\,.
\end{align}
The frequency dependence of the longitudinal susceptibility, however,
requires additional discussion.

Generally, the susceptibility, $\chi_{ij}(\omega)$, describes the
response of the magnetic impurity to a local magnetic field that oscillates
with frequency $\omega$. At low frequencies, the variation of the dissipative
part of the longitudinal component $\chi''_{zz}(\omega)$ given by
Eq.~(\ref{eq:30}) can be understood in the framework of the Debye
mechanism~\cite{Debye} of relaxational losses: At $\omega =0$, relaxation
caused by the exchange interaction between the local magnetic moment and
itinerant electrons establishes equilibrium Gibbs occupation factors for the
two Zeeman-split levels. At finite but small frequency $\omega$, the Zeeman
splitting, which is caused by the sum of a constant and a slowly-varying
magnetic field, changes with time slowly, and the relaxation acts to adjust
the occupation factors to the instant values of the Zeeman splitting. The
adjustment occurs via the emission (or absorption) of particle-hole pairs with
energy $\varepsilon_{ph} \sim \omega_Z(B)$ by flips of the local spin.  It is
the time variation of the occupation factors of the Zeeman-split levels that
leads to dissipation. In the limit $\omega\to 0$, the leading term in
$\chi''_{zz}(\omega)$, according to Eq.~(\ref{eq:30}), is
\begin{align}\label{eq:33}
\left.\chi''_{zz}(\omega)\right|_{\rm Debye}  &= \chi^D_L T_1 \omega \,.
\end{align}
As was already mentioned, in the weak-field case $1/T_1$ is given by
Eq.~(\ref{eq:15}). In the limit $\omega_Z(B)\gg T$, the time $T_1$ was
found~\cite{Goetze71} to be $T_1=T_2/2$ with $T_2$ of Eq.~(\ref{eq:32}).  The
contribution (\ref{eq:33}) to $\chi''_{zz}$ from the Debye relaxational losses
is valid at arbitrary ratio $\omega_Z(B)/T$. Note however that despite the
fact that Eq.~(\ref{eq:33}) describes dissipation at low frequency the Debye
mechanism is associated with the emission of particle-hole pairs with a
comparatively high energy $\varepsilon_{ph} \sim \omega_Z(B)$.  In the limit
$\omega_Z(B) \gg T$, the Debye mechanism thus yields only an exponentially
small contribution to dissipation,
\begin{equation}
\left.\chi''_{zz}(\omega)\right|_{\rm Debye} 
=\frac{2}{\pi}\frac{(g \mu_B)^2 \omega}{T}
\frac{\ln^2{\frac{\omega_Z(B)}{T_K}}}{\omega_Z(B)}
\exp\left[-\frac{\omega_Z(B)}{T}\right].
\label{debye}
\end{equation}
The exponential smallness of $\chi^D_L$ comes from the small probability of
the thermal occupation of the highly-excited state, corresponding to the upper
of the two Zeeman-split levels. Temporal variations in this exponentially
small quantity leads to an exponentially small contribution to
$\chi''_{zz}(\omega)$.

Under these conditions, a second contribution, originating from the low-energy
part of the spectrum, $|\epsilon|\lesssim {\rm max}[\omega,T]$, becomes
important. The processes contributing here do not involve real impurity
spin-flip processes (which are exponentially suppressed), but only virtual
transitions. The starting point is the observation that the impurity
magnetization locally polarizes the Fermi sea.  If the Zeeman splitting of the
impurity is slowly varied with a small frequency $\omega$, the magnetic
polarization of the Fermi sea will adjust itself to the instantaneous
adiabatic value of the impurity magnetization. Since the spectrum of the
particle-hole pairs is continuous this adjustment results in dissipation via
the emission of pairs with small frequency $\varepsilon_{ph} \sim \omega$,
which is in contrast to the Debye-mechanism where the emitted particle-hole
pairs carry a large energy of order the Zeeman splitting.  As shown in
Appendix~\ref{app:ShibaDerivation}, this contribution to the susceptibility
can be obtained by applying Nozi\`eres' Fermi liquid theory and is adequately
accounted for by the generalized Shiba relation Eq.~(\ref{shiba1}). Evaluating
$d\langle S^z \rangle/dB$ with the help of Eq.~(\ref{eq:28}) at $T\ll g \mu_B
B$, we find for the dissipative part of the longitudinal susceptibility
\begin{align}
\label{shiba2}
\chi''_{zz}(\omega) =\frac{\pi}{8}\frac{(g\mu_B)^2\omega}
{\omega_Z^2(B)}\frac{1}{\ln^4 \frac{\omega_Z(B)}{T_K}},\;\omega\lesssim\omega_Z(B).
\end{align}

Comparing Eq.~(\ref{shiba2}) with the result for the Debye mechanism,
we see that the strong-field asymptote Eq.~(\ref{debye}) for the
latter mechanism is important only in a narrow interval of
temperatures $\omega_Z(B)\gtrsim T\gtrsim\omega_Z(B)/6$, as for all
practical purposes $\ln\ln (\omega_Z/T_K)\approx 1$. Dispensing with
that interval, we will use for the dissipative part of the
longitudinal susceptibility Eq.~(\ref{eq:30}) with $T_1=\tau_K$ in the
case $\omega_Z(B)\lesssim T$, and Eq.~(\ref{shiba2}) in the case
$\omega_Z(B)\gg T$.

At low temperatures, $T\ll T_K$, there is little effect of the
magnetic field on $\chi''(\omega)$ for weak fields, $g\mu_BB\ll T_K$.
In the strong-field regime, $\omega_Z(B)\gg T_K\gg T$, the main
contribution to the transversal part of the dissipative susceptibility
is given by Eq.~(\ref{eq:31}) with the relaxation time $T_2$ of
Eq.~(\ref{eq:32}). The longitudinal part is described by
Eq.~(\ref{shiba2}) at $\omega\ll\omega_Z(B)$. Equation (\ref{eq:31})
adequately describes the non-monotonic behavior of
$\chi''_{+-}(\omega)$, but fails at higher frequencies; similarly the
linear dependence in $\chi''_{zz}(\omega)$ does not stretch beyond
$\pm\omega_Z(B)$. In the limit $|\omega|\gg\omega_Z(B)$, the magnetic field does not affect significantly the dissipation, and Eq.~(\ref{eq:15a}) is applicable.

\subsection{Elastic and inelastic components of electron scattering}

The coupling of the impurity spin to the low-energy degrees of freedom
of the Fermi seas will lead to a broadening and redistribution of the spectral weight
of the three delta-functions in Eq.~(\ref{eq:23}).

\subsubsection{High temperatures: $T \gg T_K$}

At high temperature, $T\gg T_K$, and weak magnetic field,
$\omega_Z(B)\ll T$, the spin polarization is weak, and the elastic
component of the scattering is small. Using Eqs.~(\ref{eq:27}) and
(\ref{eq:28}) we find
\begin{align}\label{elastic}
\lefteqn{\sigma_{\rm el}(E,\omega) = }
\\\nn & = 
\sigma_{\rm tot}(E) \frac{4}{3}
\left[1-\frac{2}{\ln(T/T_K)}\right]
\left[\frac{g \mu_B B}{4 T}\right]^2\delta(\omega).
\end{align}
The major contribution to the scattering cross-section comes from the
inelastic processes. At fields satisfying the condition $\omega_Z(B)\tau_K\gg
1$, which still belongs to the domain of weak fields $\omega_Z(B)\ll T$, the
single maximum in the $\omega$-dependence of the cross-section, see
Eq.~(\ref{eq:16}), splits into three:
\begin{align}\label{eq:35}
\sigma_{\rm inel}(E,\omega) &\approx \sigma_{\rm tot}(E) 
\\\nn \times & \frac{1}{3}
\left[\delta_{\Gamma}(\omega) + \delta_{\Gamma}(\omega-\omega_Z(B)) 
+ \delta_{\Gamma}(\omega+\omega_Z(B))\right]
\end{align} 
The broadened delta-function was defined in Eq.~(\ref{eq:16a}) with a
relaxation rate, $\Gamma$, given by the inverse Korringa time, $\Gamma =
1/\tau_K$. (We neglected a small part of the spectral weight which moved to
the elastic component of the scattering cross-section).

With the increase of the ratio $\omega_Z(B)/T$, the intensity of the
elastic scattering increases, and in the strong-field limit we find
\begin{equation}
\label{eq:36a}
\sigma_{\rm el}(E,\omega) = \sigma_{\rm tot}(E) \frac{1}{3}
\left[1-\frac{1}{\ln(g \mu_B B/T_K)}\right]
\delta(\omega).
\end{equation}
Simultaneously, the maximum of $\sigma_{\rm inel}(E,\omega)$ at
negative $\omega$ gets suppressed, and the structure at
$|\omega|\ll\omega_Z(B)$ broadens and becomes asymmetric. In the limit
$\omega_Z(B)/T\gg 1$, only a single maximum at positive
$\omega$ remains in the inelastic cross-section,
\begin{eqnarray}
\sigma_{\rm inel} (E,\omega)
&=&\sigma_{\rm tot}(E) \frac{2}{3 \pi}
\frac{1}{1-e^{-\omega/T}}\frac{1}{\omega_Z(B)}
\nonumber\\
&\times&\frac{\omega/T_2}{[\omega-\omega_Z(B)]^2 + (1/T_2)^2}.
\label{eq:36}
\end{eqnarray}
Here the relaxation time $T_2$ is defined by Eq.~(\ref{eq:32}). This main
contribution to the inelastic scattering is proportional to
$\chi''_{+-}(\omega)$ and comes from the spin-flip processes. The comparison
of Eqs.~(\ref{debye}) and (\ref{shiba2}) with Eq.~(\ref{eq:31}) shows that at
$\omega_Z(B) \gg T_K$ the effect of the dissipative part of longitudinal
susceptibility is small starting from $\omega_Z(B)/T\gtrsim 4$. Under this
condition, $\chi''_{zz}(\omega)$ yields a contribution to $\sigma(E,\omega)$
which is small compared to Eq.~(\ref{eq:36}).

The high-frequency tail, $|\omega| \gg
{\rm max}[T_K, g\mu_B B, T]$, is unaffected by the Zeeman splitting and still given by Eq.~(\ref{eq:17}).

\subsubsection{Low temperatures: $T \ll T_K$}

\begin{figure}
\includegraphics[width= 0.9\linewidth]{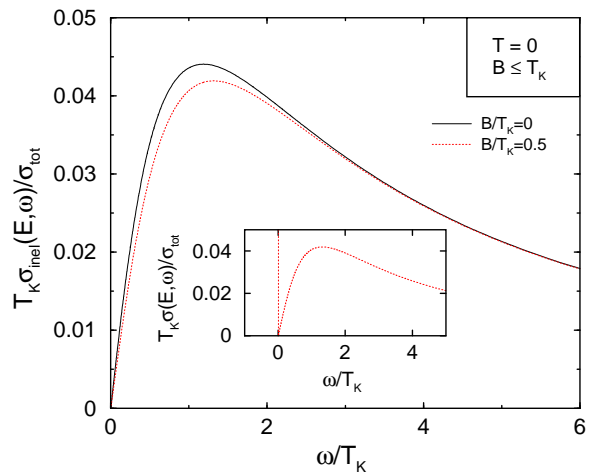}
\caption{\label{fig:2b} (Color online) NRG result for $\sigma(E,\omega)$ at $T=0$ for
magnetic fields $g \mu_B B < T_K$. The difference of the curves indicate the
scattering weight which for $B>0$ is transfered from the inelastic to the
elastic component leading to a delta-function peak at $\omega=0$, as sketched
in the inset.  }
\end{figure}

We turn now to the opposite limit of small temperature, $T\ll T_K$. At
weak magnetic field, $g\mu_BB \lesssim T_K$, the low-frequency behavior of
the scattering cross-section is beyond perturbation theory.  In this
regime, the electron scatters from a fully developed, many-body Kondo
singlet. Here we can use the Shiba relation Eq.~(\ref{eq:20}) to
access the low-frequency tail of the cross-section. In the presence of
a magnetic field there are additional corrections to the Shiba
relation of order $\mathcal{O}(\omega (g \mu_B B)^2/T_K^2)$ which are
sub-leading and are neglected in the following. We get for the
low-frequency part $|\omega| \ll T_K$
\begin{align}\label{eq:34}
\sigma(E,\omega) &= \sigma_{\rm tot}(E)
\frac{W^2}{2}
\frac{1}{1-e^{-\omega/T}} 
\\\nn &\times 
\left[\frac{1}{6}\left(\frac{g \mu_B B}{T_K}\right)^2 \delta(\omega) +
\frac{\omega}{T_K^2}\right].
\end{align}
where $W$ is again Wilson's number~\cite{Hewson}.  The scattering
cross-section decreases linearly with frequency.  At $\omega \lesssim
T$ the linear decrease crosses over into an exponential tail which
extends to negative frequencies.  In Fig.~\ref{fig:2b} NRG results at
$T=0$ for the inelastic cross-section at small magnetic fields are
compared with the NRG data at $B=0$. In finite field the slope in the linear
low-frequency regime is reduced. The difference in slope is of order
$\mathcal{O}(g \mu_B B/T_K)^2$, a correction alluded to but neglected
in Eq.~(\ref{eq:34}). This difference however accounts for the
reduction of the inelastic scattering weight. The weight of order
$\mathcal{O}(g \mu_B B/T_K)^2$ is transfered from the inelastic to the
elastic scattering contribution leading to a delta peak at $\omega=0$,
as sketched in the inset of Fig.~\ref{fig:2b}. In contrast to the case
of high temperatures ($T \gg T_K, g \mu_B B$) the elastic scattering
contribution now does not sit on top of a large Lorentz-peak but is
rather located within the scattering pseudogap. Although its weight is
small, here it is  easily distinguishable from the background.  The
crossover from the linear dependence on $\omega$ to the high-frequency
behavior occurs at $\omega\sim T_K$, where the inelastic scattering
cross-section has a maximum. The high-frequency tail is still given by
the perturbative expression (\ref{eq:17}).  

\begin{figure}
\includegraphics[width= 0.9\linewidth]{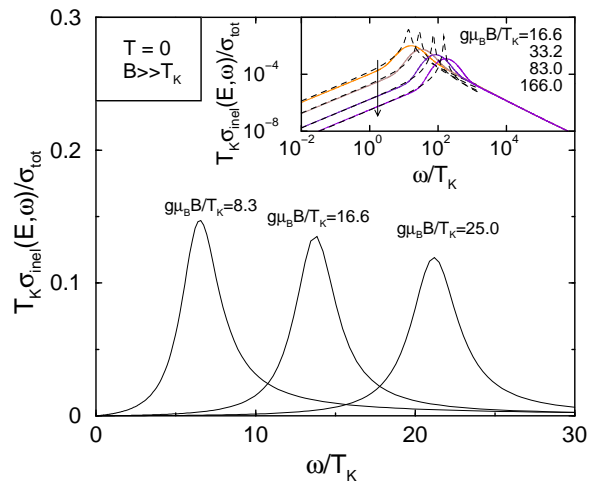}
\caption{\label{fig:2c} (Color online) Inelastic scattering cross-section for several
magnetic field values, $B \gg T_K$, at $T=0$ according to Eq.~(\ref{eq:36}),
which is applicable unless $\omega \gg B$. Note that the position of the
resonance is shifted away from the Zeeman energy as described by
Eq.~(\ref{eq:25}). The inset shows the comparison of the analytical result
(\ref{eq:36}) (dotted lines) with NRG calculation (colored lines). The
low-frequency tail matches nicely. However, the NRG overestimates the width of
the peak.}
\end{figure}

When the magnetic field is increased above the Kondo temperature,
$g \mu_B B \gg T_K$, the elastic and inelastic components of the scattering
cross-section are given by Eqs.~(\ref{eq:36a}) and (\ref{eq:36}),
respectively.  The elastic peak at $\omega=0$ now exhausts almost the full
spectral weight of the longitudinal correlator, {\it i.e.}~it accounts for
approximately $1/3$ of the total scattering cross-section, see Fig.~\ref{fig:3}. The remaining $2/3$
of the total spectral weight are to be found in the extended structure of the
Zeeman satellite (\ref{eq:36}) centered at $\omega=\omega_Z(B)$.
The effect of Zeeman spitting on the cross-section is confined to the
region of energies $|\omega|\lesssim\omega_Z(B)$. At $|\omega| \gg
{\rm max}[T_K, g\mu_BB, T]$ the behavior of $\sigma (E,\omega)$ is
again given by Eq.~(\ref{eq:17}).
 
In Fig.~\ref{fig:2c} the inelastic cross-section is shown in the limit
of large magnetic fields, $g \mu_B B \gg T_K$, as given by
Eq.~(\ref{eq:36}). The inset compares the result with the NRG. The
low-frequency and high-frequency asymptotes are reproduced in the
numerical calculation fairly well. The deviation in the width of the
Zeeman peak, however, demonstrates the limitation of the NRG method.
Due to the logarithmic frequency resolution the NRG tends to
overbroaden any peak in the spectral function centered around a {\em
  nonzero} frequency.

\section{Possible Experiments}
\label{sec:Experiments}

As we have shown above, the differential scattering cross-section of the magnetic impurity 
shows a rich structure in frequency space. In the following we suggest two experiments that are sensitive to the dynamics of a Kondo impurity and from which, in principle, the energy-resolved scattering cross-section can be extracted.

\subsection{Mesoscopic wires}

\begin{figure}
\includegraphics[width= 0.9\linewidth]{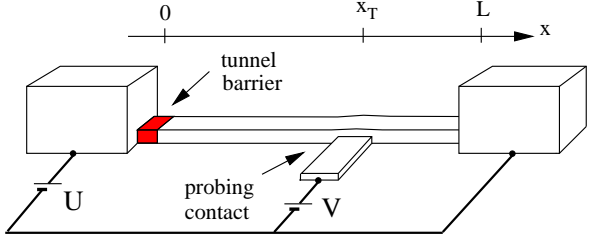}
\caption{\label{fig:PothierExp} (Color online) Experimental setup of Pothier
  {\it et al.}\cite{Pothier} with an additional tunnel barrier which limits
  the injection of hot electrons into the wire. The voltage drop across the
  wire is denoted by $U$. The voltage applied at the probing contact is $V$.
  }
\end{figure}

Inelastic scattering off magnetic impurities has been identified to be at the
origin of an anomalously large energy relaxation in mesoscopic metallic
wires~\cite{Pothier}. 
We propose a modification of the original experiment performed by Pothier {\it et al.}~\cite{Pothier}
that allows in principle to access the scattering cross-section considered in
this paper. We assume that the wire is connected to the reservoirs on one end by
an open contact and on the other via a tunnel junction, see Fig.~\ref{fig:PothierExp}. In the limit of small
transparency of the tunnel junction the wire is almost in equilibrium; only a
small amount of energetic quasi-particles tunnel into the wire and relax their
energy during scattering processes on magnetic impurities. 
In lowest order in the transparency of the tunnel junction we can treat this 
relaxation mechanism in terms of the differential scattering 
cross-section, $\sigma(E,\omega)$ of a test particle coming with
energy $E$ in an otherwise equilibrium system. 

Consider a mesoscopic wire of length $L$. The equilibrium
distribution in the right and left reservoir is given
by a Fermi function, $f_{\rm F}(E)$ and $f_{\rm F}(E - e U)$,
respectively, where the energy $E$ is measured with respect to the
chemical potential of the right reservoir. The voltage drop across the
wire is $U$. 
Within the wire the distribution function, $f(E; x,U)$, will depend on 
the position across the wire, $x \in [0,L]$. 
It is determined
by the relaxation mechanisms and carries information on the
differential scattering cross-section $\sigma(E,\omega)$. This
distribution function is probed by an additional tunnel contact that
is attached to the wire at a certain position $x_T \in [ 0,L ]$ and
connects it to a conductor with a sharp feature in the density of
states. Measurement of a small tunneling current through this
auxiliary contact as a function of voltage $V$, see
Fig.~\ref{fig:PothierExp}, allows one~\cite{Pothier} to probe the
electron energy distribution. This way, the distribution function
$f(E;x_T,U)$ in the wire at some point $x_T$ in the presence of a bias $U$
applied across the wire was investigated~\cite{Pothier,Anthore}. The
sharp feature in the electron density of states in the probe was due
to its superconducting state~\cite{Pothier} (the BCS anomaly) or due to
the Coulomb interaction in a low-dimensional diffusive electron
system~\cite{Anthore} (zero-bias anomaly). In the following we show
that measurement of the derivative $\partial f(E;x_T,U)/\partial U$ in
a modified (compared to Ref.~\onlinecite{Pothier}) setup of
Fig.~\ref{fig:PothierExp} allows one to access the inelastic scattering
cross-section $\sigma(E,\omega)$.

The distribution function within the wire
is governed by the diffusive Boltzmann equation~\cite{Nagaev}
\begin{align} \label{DiffusiveBoltzman}
- D \frac{\partial^2 f(E; x, U)}{\partial x^2} = I[f]
\end{align}
where $D$ is the diffusion coefficient of the wire. The collision
integral is local in space,
\begin{align}
I[f] &= c_{\rm imp} v_F 
\\\nn& \times \int_{-\infty}^\infty d\omega \left[
f(E) \left(1-f(E-\omega)\right) \sigma(E,\omega) 
\right.\\\nn
&- \left.
\left(1-f(E)\right) f(E-\omega) \sigma(E-\omega,-\omega)
\right]\,.
\end{align}
where $c_{\rm imp}$ is the impurity concentration within the wire and
$\sigma(E,\omega)$ is the differential cross-section of a single
magnetic impurity, and for notational convenience the dependence of the 
distribution function, $f$, on $x$ and $U$ has been omitted.
The boundary condition at the open contact to the right reservoir is
simply $f(E; x = L,U) = f_{\rm F}(E)$. The boundary condition at the tunnel
contact, which connects the wire to the left reservoir, is determined by
current conservation
\begin{align} \label{BoundaryCondition-1}
g_T \left(f_{\rm F}(E - eU) - f(E; x=0, U)\right) = \hspace{5em}
\\\nonumber
= - \nu D \frac{\partial f(E; x=0, U)}{\partial x}
\end{align}
where $g_T$ is the dimensionless conductance of the tunneling contact
and $\nu$ is the density of states of the wire.

In zeroth order in the collision integral we obtain the solution
\begin{align}
f^{(0)}(E; x, U) = f_{\rm F}(E) \frac{L_0+x}{L_0+L} + f_{\rm F}(E-eU) \frac{L-x}{L_0+L}\,,
\end{align}
where we introduced the length $L_0$; the relation of $L_0$ to the
length of the wire, $L$, is determined by the ratio of conductances of
the wire and the tunneling contact, $L_0/L = g_w/g_T$, with $g_w = \nu
D/L$. In the limit of large transparency of the tunneling contact,
$L_0/L \ll 1$, the obtained solution reduces to the well-known formula
for the distribution function of a diffusive wire with open
contacts~\cite{Nagaev}.  However, we are focusing on the other limit
of a large tunneling barrier, $L/L_0 \ll 1$, where we get
$f^{(0)}(E; x,U) = f_{\rm F}(E) + \delta f^{(0)}(E; x,U)$ with
\begin{align} \label{ZerothOrderFunction}
\delta f^{(0)}(E; x,U) = \hspace{14em} 
\\\nonumber 
= \left( f_{\rm F}(E-e U) - f_{\rm F}(E)\right) \frac{L-x}{L_0} + 
\mathcal{O}\left(\frac{L}{L_0}\right)^2\,.
\end{align}
The deviation of the energy distribution in the wire from the one in
the right reservoir is of first order in the small parameter $L/L_0$.

In the following we consider the correction to the distribution function in 
lowest order in the collision integral and in the small parameter $L/L_0$.
We get in leading order in $L/L_0$
\begin{align}
I[f^{(0)}] &= c_{\rm imp} v_F\,\frac{L-x}{L_0} \left( f_{\rm F}(E-e U) - f_{\rm F}(E)\right)
\\\nn &\times
\int_{-\infty}^\infty d\omega\, \sigma(E,\omega) 
\left(1-e^{-\beta \omega}\right)
\\\nn &\qquad\qquad
\left(f_{\rm F}(E-\omega-e U) - f_{\rm F}(E-\omega)\right)\,.
\end{align} 
Returning now to Eq.~(\ref{DiffusiveBoltzman}), we are able to find the
correction to the distribution function. The energy dependence of $\partial
f(E; x, U)/\partial U$ within the interval $0 < E < eU$ is caused by
electron energy relaxation. At $T=0$ it is given by
\begin{align} \label{sigmaprobe}
\frac{\partial f(E; x, U)}{\partial (e U)} = \hspace{15em} 
\\\nonumber 
= \frac{c_{\rm imp} v_F}{D L_0} \left(-\frac{x^3}{6}+\frac{L x^2}{2} - \frac{L^2 x}{3} \right)\sigma(eU,eU-E)\,.
\end{align}
The structure of the distribution function in this energy interval is
directly related to the differential scattering cross-section
$\sigma(E,\omega)$. The simple relation between $\partial f (E; x,
U)/\partial U$ and the cross-section holds as long as the events of
scattering off magnetic impurities occur rarely over the time limited
by the diffusion of an electron across the wire.  Note however that
in addition to Eq.~(\ref{sigmaprobe}), there is a sharp
contribution at the edge of the energy interval, $E=eU$, resulting
from the zeroth order contribution (\ref{ZerothOrderFunction}) to the
non-equilibrium distribution function. At finite temperature, this
limits the experimental accessibility of $\sigma(E,\omega)$ for
$\omega \lesssim T$.

\subsection{Quantum dot in the Kondo regime}

\begin{figure}
\includegraphics[width= 0.8\linewidth]{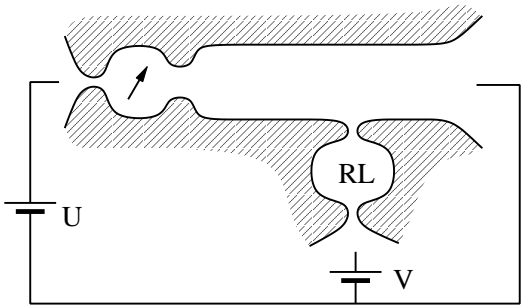}
\caption{\label{fig:QuantenDotExp} Experimental quantum dot setup. A wire
leads to a quantum dot in the Kondo regime (indicated by the arrow), which is
in addition weakly connected to another reservoir at a voltage $U$. The
resonance--level (RL) quantum dot acts as a probe.  }
\end{figure}

A second experimental possibility is very similar in spirit to the first one
but considers a quantum dot setup. The starting point is a semiconductor-based
ballistic wire that has on its right hand side contact with a large reservoir,
see Fig.~\ref{fig:QuantenDotExp}. On the other end, the wire is connected to a
quantum dot in the Kondo regime. In this regime, the spin of the dot forms a
many-body ground state with the electrons in the wire. Electrons injected at a
bias $U$ into the wire through the quantum dot form a non-equilibrium
distribution, which is probed via an auxiliary weak contact having potential
$V$.  The auxiliary contact consists of a second quantum dot, marked RL in
Fig.~\ref{fig:QuantenDotExp}, which is tuned to the resonant tunneling
regime. In the case of a sharp resonance, the setup of
Fig.~(\ref{fig:QuantenDotExp}) allows one to measure the electron energy
distribution in the quantum wire. This non-equilibrium distribution, in turn,
is sensitive to the inelastic transport through the Kondo dot and bears
signatures of the differential scattering cross-section of the Kondo spin.

If the left reservoir is disconnected from the Kondo dot the electrons
within the ballistic wire have a equilibrium Fermi distribution,
$f^{(0)}(E) = f_F(E)$. The injection of hot electrons from the left
reservoir will lead to a non-equilibrium correction to the
distribution function of the right-movers within the wire, $f(E) =
f_F(E) + \delta f(E)$. We obtain in lowest order in tunneling between
the left lead and the dot:
\begin{align} 
v_F \delta f(E) =
v_F\!\! \int d\xi 
\sigma(\xi,\xi-E) \left(f_F(\xi-e U) (1-f_F(E))
\right.
\nn\\ \left. 
- e^{-\beta (\xi-E)} f_F(E) (1-f_F(\xi-e U))\right),
\end{align}
where we used already the detailed balance relation,
$\sigma(E,\omega) e^{-\beta \omega} = \sigma(E-\omega, -\omega)$.
After taking the derivative with respect to $U$, the above equation
simplifies considerably at $T=0$, and we get
\begin{align} 
\frac{\partial f(E)}{\partial(e U)} = \sigma(eU, eU - E)\,.
\end{align}
The measurement of this quantity with help of the auxiliary contact
thus yields direct access to the differential inelastic scattering
cross-section of a Kondo system.

\section{Summary} 

We analyzed inelastic scattering of energetic electrons off a magnetic
impurity. For such scattering, the dependence of the differential
cross-section, $\sigma(E,\omega)$, on energy $E$ of the incoming electron is
logarithmically weak at $E\gg T_K$, and arises from the renormalization of the
exchange coupling. In the leading logarithmic approximation, the total
cross-section $\sigma_{\rm tot}=\int d\omega\sigma (E,\omega)$ is proportional
to $1/\ln^2(E/T_K)$, in agreement with Ref.~\onlinecite{Abrikosov1965}. More
interestingly, the electron scattering is inelastic, and the dependence of
$\sigma(E,\omega)$ on the energy transfer, $\omega$, is determined by the
spin-spin correlation function of the impurity or, equivalently, by the
dissipative part of the impurity spin susceptibility, $\chi''$.  In the
absence of magnetic field, the elastic component of scattering appears only in
order $1/\ln^4(E/T_K)$.

Our findings confirm and quantify the conclusion of Ref.~\onlinecite{Natan}
regarding the inelastic nature of Kondo scattering, and also provide a clear
physical picture of the mechanism of inelastic scattering. In the absence of
magnetic field, the inelastic scattering cross-section is parametrically
larger than the elastic one. The typical energy transfer $|\omega|$ in an
inelastic scattering event is however small compared to $E$.  At high
temperatures, $T\gg T_K$, the characteristic energy transfer is determined by
the Korringa relaxation rate of the magnetic impurity, and at low
temperatures, it is defined by the value of $T_K$. In the high-temperature
limit, the cross-section is maximal at $\omega=0$, and at $T\ll T_K$ it
reaches its maximum at $\omega\sim T_K$.  The decrease of the cross-section in
the domain $\omega\gg T_K$ is remarkably slow, $\sigma(E,\omega) \propto
[\omega \ln^2 (\omega/T_K)]^{-1}$. The domain of intermediate energy
transfers, $\omega\sim T_K$, is covered by NRG calculations. The numerical
results fit well with the analytically evaluated asymptotes at $\omega\ll T_K$
and $\omega\gg T_K$. In the presence of an external magnetic field, the Zeeman
splitting of the magnetic impurity levels results in the appearance of an
elastic component of electron scattering already in the leading logarithmic
order (in $E/T_K$).
 
Finally, we proposed possible hot-electron experiments with a metallic
mesoscopic wire and with a semiconductor quantum-dot device which in
principle allow one to access the differential scattering
cross-section of a localized magnetic moment.

\acknowledgments This work was supported by NSF Grants DMR02-37296 and
EIA02-10736 (L.G.), by the Deutsche Forschungsgemeinschaft (DFG) grant GA
1072/1-1 (M.G.), the DFG "Center for Functional Nanostructures" (P.W.), DFG
SFB631 and the European "Spintronics" RTN (J.v.D.), and by Hungarian Grants
OTKA D048665, T048782, T046303 (L.B.). L.B. is a grantee of the Bolyai Janos
Scholarship.

\appendix

\section{Derivation of Shiba relation}
\label{app:ShibaDerivation}

Here we provide a simple derivation of Shiba relation~\cite{Shiba}, using
Nozi\`eres' idea of a low-temperature Fermi liquid description of the
Kondo problem.

Within Nozi\`eres' theory, at $T \ll T_K$ the effect of a weak ($g\mu_BB\ll T_K$)
magnetic field applied to a Kondo impurity is described by a
local-field Hamiltonian,
\begin{equation}
H_B=\frac{2 \chi_0}{g\mu_B\nu} B
\sum_{k,k',\sigma} s^z_{\sigma\sigma}\psi_{k\sigma}^\dagger\psi_{k'\sigma}.
\label{eq:a1}
\end{equation}
Here $\nu$ is the density of states at the Fermi level,
$\chi_0=[W(g\mu_B)^2]/(4T_K)$ is the linear susceptibility,
summation over $k$ and $k'$ occurs within a shell of states $\Delta k$
sufficiently close to the Fermi level ($\Delta k\sim T_K/v_F$ with
$v_F$ being the Fermi velocity), and field $B$ is applied along the $z$-axis. 
One may easily check that the action of the field described by the
Hamiltonian (\ref{eq:a1}) indeed results in a local magnetization
$M=\chi_0B$. For that one starts with the evaluation of the
spin-dependent scattering phase $\delta_\sigma$ off the local
perturbation Eq.~(\ref{eq:a1}) using the Born approximation,
\begin{equation}
\delta_\sigma=\pi\sigma\nu\cdot\frac{\chi_0}{g\mu_B\nu}B,\;\quad\sigma=\pm 1.
\label{eq:a2}
\end{equation}
Having the phase difference $\delta_+-\delta_-$, we evaluate the
magnetization using the Friedel sum rule,
\begin{equation}
M=\frac{g\mu_B}{2}\frac{\delta_+-\delta_-}{\pi}=\chi_0B.
\label{eq:a3}
\end{equation}

Having the right form of the local perturbation, we now allow for a slow
variation of the field, $B=B_0\cos(\omega t)$, assuming that the
frequency $\omega\ll T_K$. Next we evaluate the energy absorption rate $w$
caused by such time-dependent perturbation. Using the Fermi Golden
rule, we arrive at
\begin{align}
w &= \pi \omega \left[\frac{\chi_0 B_0}{g\mu_B\nu}\right]^2 
\nu^2\int
d\epsilon f(\epsilon)[f(\epsilon-\omega)-f(\epsilon+\omega)]
\nonumber\\
&=\pi\omega^2 \left[\frac{\chi_0 B_0}{g\mu_B}\right]^2.
\label{eq:a4}
\end{align}
In the last line, we discarded corrections of order $\mathcal{O}(e^{-T_K/T})$ 
arising from the boundaries of the energy integral. 
Recalling finally that $w=\frac{1}{2}\omega\chi''(\omega)B_0^2$, we arrive at the
Shiba relation Eq.~(\ref{shiba}). 

Using the framework of the above derivation, it is straightforward to
generalize the Shiba relation to the case of a weak slowly varying 
field applied to the local moment on top of a
time-independent field $B$ of arbitrary strength. In the
generalized relation, $\chi_0$ is the static {\sl differential}
susceptibility, and the relation is applicable in the regime $\omega,
T\ll {\rm max} \{B, T_K\}$. 

\begin{figure}[tbp]
\includegraphics[width= 0.9\linewidth]{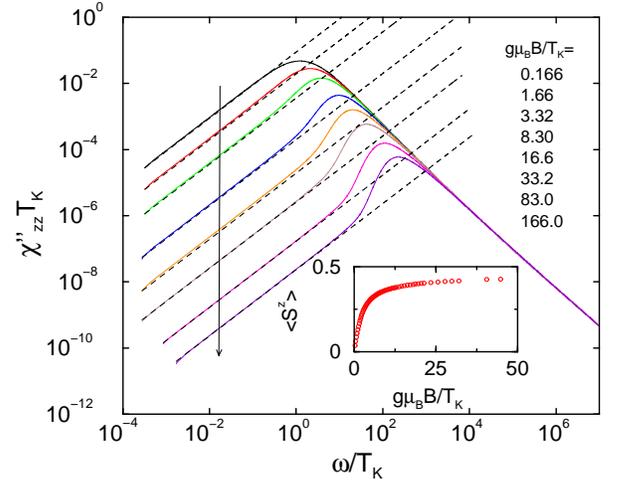}
\caption{(Color online) NRG comparison of the low-frequency asymptote of $\chi''_{zz}$ with
the prediction of the generalized Shiba relation Eq.~(\ref{shiba1}) (dotted
lines). The inset shows the numerically evaluated magnetization whose
derivative enters Eq.~(\ref{shiba1}).}
\label{fig:Shiba}
\end{figure}

We assume that the basis of the effective low-energy Hamiltonian has been
chosen such that it incorporates already the effect of the time-independent
local magnetic field $B$. Consider now a small perturbation to this effective
Hamiltonian induced by a small change in the applied local magnetic field $B
+\delta B$,
\begin{equation}
H_{\delta B} =\frac{2}{g\mu_B\nu}\frac{\partial M}{\partial B}\delta B
\sum_{k,k',\sigma} s^z_{\sigma\sigma} c_{k\sigma}^\dagger c_{k'\sigma}.
\label{eq:a5}
\end{equation}
The summation over $k$ and $k'$ is bounded by $|k|,|k'| \lesssim {\rm
  max}\{g \mu_B B,T_K\}/v_F$.  The prefactor can be determined in the
same way as before. In contrast to the limit $B=0$, here the resulting
phase shift yields information about the change in magnetization
$M(B+\delta B) - M(B) = (\partial M / \partial B)\delta B$, where
$\partial M / \partial B$ is the differential susceptibility. The same
arguments as above will yield the generalized Shiba relation
\begin{equation}
\chi''_{zz}(\omega) = 
2 \pi \omega \left[\frac{\partial \langle S^z \rangle}{\partial B}\right]^2
\label{shiba1}
\end{equation}
Here $\langle S^z\rangle$ is the equilibrium average spin value in the
presence of field $B$. In the perturbative regime, the average is
given in Eq.~(\ref{eq:28}).

In Fig.~\ref{fig:Shiba} the prediction of the generalized Shiba relation
(\ref{shiba1}) is illustrated with the NRG result. The susceptibility,
$\chi''_{zz}$, and the non-linear static susceptibility, $\chi_0 = g \mu_B
\partial \langle S^z \rangle/\partial B$, have been independently evaluated
with the NRG. The dashed line in Fig.~\ref{fig:Shiba} is plotted with the help
of Eq.~(\ref{shiba1}) and compares well with the low-frequency asymptote of
$\chi''_{zz}$.

\section{RG equation for the impurity $g$-factor}
\label{app:RG-gfactor}

\begin{figure}
\centering
(a)
\includegraphics[width= 5em]{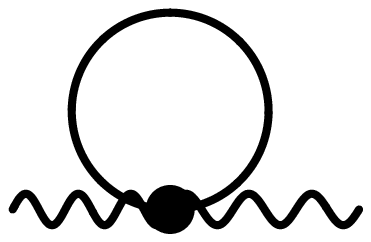}
\quad
(b)
\includegraphics[width= 7em]{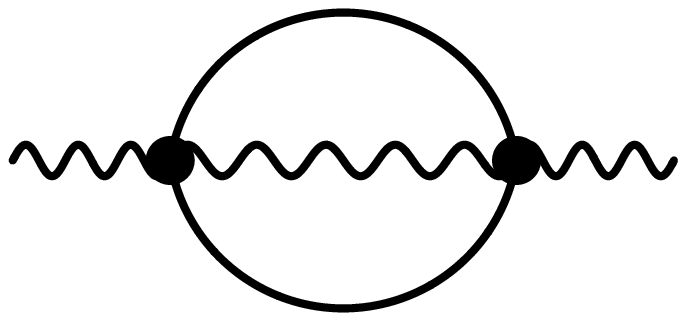}
\quad
(c)
\includegraphics[width= 7em]{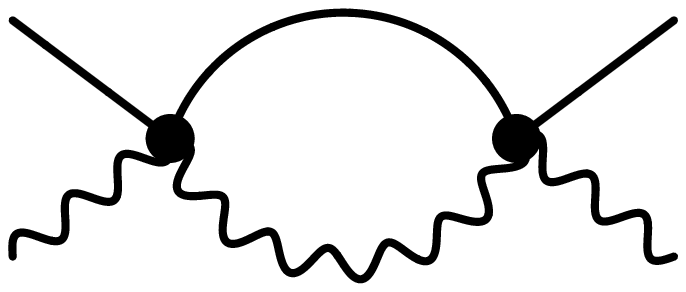}
\caption{\label{fig:Diagrams} (a) First order Knight-shift diagram; (b)
two--loop self-energy correction; (c) one--loop vertex correction. The wiggled
line represents the propagator of the Abrikosov fermions and the solid line
the electron propagator. The dot signifies the Kondo interaction.}
\end{figure}

We present a derivation of the two-loop RG equation for the impurity
$g$-factor of the Kondo model and, in particular, explain the different roles
played by the impurity and conduction electron $g$-factor in the
renormalization process. To this end, we will use Abrikosov's pseudo-fermion
representation\cite{Abrikosov1965} for the impurity spin, ${\bf S} = f^\dagger
\frac{1}{2} \boldsymbol{\sigma} f$, where $f^\dagger = (f^\dagger_\uparrow,
f^\dagger_\downarrow)$ in a compact spinor notation and $\boldsymbol{\sigma}$
is the vector of Pauli matrices. We will need the action of the Kondo model,
which consists of three parts, $\mathcal{S} = \mathcal{S}_s + \mathcal{S}_d
+\mathcal{S}_K$.  The quadratic part of the Abrikosov pseudo-fermions reads
\begin{align} \label{ActionPseudoFermions}
\mathcal{S}_d &=
\int_0^\beta d \tau f^\dagger(\tau) 
\left[\partial_\tau -\lambda_0 - g \mu_B\, \frac{1}{2} \boldsymbol{\sigma}^a B_a \right] f(\tau)\,,
\end{align}
where $g$ is the impurity $g$-factor and $B^a$ is the magnetic field, which is
taken to point in the z-direction $B^a = B\, \delta_{az}$. In order to enforce
the Hilbert space constraint, $f^\dagger f = 1$, a chemical potential,
$\lambda_0 \to \infty$, is introduced~\cite{Abrikosov1965}. The Kondo
interaction is given by
\begin{align}
\mathcal{S}_K &=
\int_0^\beta d \tau 
\left(\Psi^\dag(\tau) \frac{1}{2} \boldsymbol{\sigma}^a
\Psi(\tau)\right) J_{ab}
\left(f^\dagger(\tau) \frac{1}{2} \boldsymbol{\sigma}^b f(\tau)\right) ,
\end{align}
where the local electron operator at the impurity site is $\Psi^\dagger = \int
\frac{d k}{2\pi v_{\rm F}} (c^\dagger_{k\uparrow}, c^\dagger_{k
\downarrow})$. We allow for different values of the exchange interaction in
the direction orthogonal and perpendicular to the magnetic field, $(J_{ab}) =
$diag$\{J_\perp,J_\perp,J_\parallel\}$. Finally, the quadratic part of the
s-electrons reads
\begin{align} \label{ActionElectons}
\mathcal{S}_s = \int_0^\beta d\tau 
\int\limits_{-D}^{D} \frac{dk}{2\pi v_{\rm F}}
c^\dag_{k \sigma}(\tau) 
\left[\partial_\tau + k \right] 
c^\pdag_{k \sigma}(\tau) \,.
\end{align}
In the presence of a Zeeman energy for the s-electrons, the Fermi sea of the
spin-up and -down electrons are shifted with respect to each other giving rise
to a finite Pauli magnetization. In Eq.~(\ref{ActionElectons}) we assumed that
the band has already been symmetrized around the respective Fermi energies by
integrating out a finite number of electronic degrees of freedom. This process
results in a perturbative renormalization of the impurity $g$-factor due to
the so-called Knight shift. The first-order Knight-shift diagram is shown in
Fig.~\ref{fig:Diagrams}a.  The $g$-factor, $g$, appearing in
(\ref{ActionPseudoFermions}) is therefore understood to be already the
Knight-shifted impurity $g$-factor,
\begin{align} \label{KnightShift}
g = g_i - \frac{J_\parallel\nu}{2} g_e + \mathcal{O}(J_\parallel\nu)^2\,,
\end{align}
where $g_i$ and $g_e$ are the bare impurity- and electronic $g$-factors,
respectively, and the density of states is $\nu = 1/(2\pi v_{\rm F})$. As is
clear from Eq.~(\ref{KnightShift}) the electronic $g_e$ can be absorbed in an
effective impurity $g$-factor. The Knight shift is thus only a perturbative
phenomenon and, in particular, is not enhanced by logarithmic
renormalizations. This is expected since the Pauli magnetization affects only
electronic states far away from the Fermi edge deep inside the Fermi sea.

The field theory can be renormalized~\cite{ZinnJustin} with a wave-function,
impurity $g$-factor and Kondo-coupling renormalization (in addition to a
counterterm absorbing a shift in the unphysical chemical potential
$\lambda_0$),
\begin{align}
f = \sqrt{Z}  f^R\,,
\quad
g = \frac{g^R}{Z}\,,
\quad
J_{ab} = 
\frac{J^R_{ab}}{Z}\,.
\end{align}
We compute the renormalization 
of the Kondo coupling to one-loop and the
renormalization of the wave-function and $g$-factor to two-loop
order. The corresponding diagrams are shown in Fig.~\ref{fig:Diagrams}b and c.
The resulting RG equations for the Kondo vertex are the well-known poor man's
scaling equations~\cite{Anderson}
\begin{align} \label{PoorMansScaling}
\frac{d (J_\perp \nu)}{d \ln D} &= - (J_\perp\nu) (J_\parallel\nu)\,,
\qquad
\frac{d (J_\parallel \nu)}{d \ln D} = - (J_\perp \nu)^2\,.
\end{align}
For the wave-function renormalization we obtain,
\begin{align}
\frac{d \ln Z}{d \ln D} &= \frac{1}{8} \left((J_\parallel\nu)^2 + 2 (J_\perp \nu)^2\right)\,.
\end{align}
Finally, the main result is the RG equation for the $g$-factor,
\begin{align} \label{RG-gfactor}
\frac{d g}{d \ln D} &= \frac{1}{2} g\, (J_\perp \nu)^2\,.
\end{align}
Solving this equation in the isotropic case, $J_\perp =
J_\parallel = J$, and expanding the result in leading logarithmic order we get
\begin{align}
g(D) = g_i \left(1 - \frac{1}{2 \ln \frac{D}{T_K}} + 
\left(1 - \frac{g_e}{g_i} \right) \frac{J \nu}{2} \right)
\end{align}
where we already substituted the Knight-shifted $g$-factor
(\ref{KnightShift}). In the scaling limit, $J \to 0$ while $T_K$ is held
fixed, any dependence on the electronic $g$-factor, $g_e$, vanishes, and we
obtain the result cited in the body of the paper, Eq.~(\ref{eq:24}). In
particular, note that in the absence of a Knight shift, $g_e = 0$, the
perturbative correction to the $g$-factor starts only in second order in the
exchange coupling $J$ but, nevertheless, after renormalization group
improvement leads to a correction which is of leading logarithmic order.  At
zero temperature the RG equation for the $g$-factor coincides with the RG
equation for the impurity magnetization, which was already determined in
Ref.~\onlinecite{Abrikosov70}.


\begin{thebibliography}{}
\bibitem{kroha1} Some extensions of the theory beyond the
  leading-logarithmic approximation were suggested in
  Ref.~\protect\onlinecite{Grabert} and in
  Ref.~\protect\onlinecite{kroha2}. The goal of these extensions was
  to adapt the theory to smaller energies $E$. However, at $E\gtrsim
  T_K$, multi-particle scattering processes become important, and the
  quantity $K$ is poorly defined; this raises concerns about the
  validity of the suggested extensions.
\end{thebibliography}
\end{document}